# Tunable plasmonic properties of spatially overlapping asymmetric nanoparticle dimers

Merneh Mandado Mana[1, 2], Bereket Dalga Dana[3], Alemayehu Nana Koya[2,4*], Boyu Ji[1*], Jingquan Lin[1*]

1. School of Science, Changchun University of Science and Technology, Changchun 130022, China

2. Wolaita Sodo University College of Natural and Computational Sciences department of Physics, P.O.Box 138, Ethiopia

3. Jinka University, College of Natural and Computational Sciences, department of Physics, P. O. Box 165, Ethiopia

4. GPL Photonics Laboratory, State Key Laboratory of Luminescence and Applications, Changchun Institute of Optics, Fine Mechanics and Physics, Chinese Academy of Sciences, Changchun 130033, China

* alemayehu.koya@gmail.com; *jiboyu@cust.edu.cn; * linjingquan@cust.edu.cn

**Abstract**

In this work, the plasmonic properties of nanoparticle dimers with optical responses over a wide spectral range have been investigated by varying the inter-particle gap, dimer geometry, gap morphology, nanoparticle composition, and refractive index of the surrounding medium. In particular, we have theoretically investigated the plasmonic properties of spatially overlapping symmetric gold nanodisks, shape-asymmetric gold nanodisk nanoplates, and compositionally asymmetric gold-silver nanodisk dimers by varying the gap separation from touching to overlapping regime. In such a configuration, we have observed the appearance of a dominant bonding dimer plasmon (BDP) mode that blue-shifts as gap separation turns from touching to overlapping. In addition, it is found that asymmetric dimer produces a broader resonance shift compared to symmetric dimer because of the hybridization of bright and dark plasmon modes, making it a viable option for sensing applications. It is also found that blue shifting of the plasmon mode occurred by changing the gap morphology of the contacting region of the dimer for fixed nanoparticle size and dimer overlapping. Moreover, we explored the influence of overlapping nanoparticle dimer thickness and observed a notable resonance shift by varying the thickness of the nanoparticle dimer. Finally, based on this tunable resonance shift, we explored the sensing applications of bonding dimer plasmon mode with optimized geometries. Thus, the computed figure of merit (FOM) of the overlapping symmetric, shape-asymmetric, and compositionally asymmetric nanoparticle dimers were found to be 1.55, 2.08, and 3.04, respectively, and comparative advantages among the three configurations with implications for surface-based sensing have been thoroughly discussed.

1. **Introduction**

Recently, metallic nanoparticle dimers consisting of two metal nanoparticles separated by nanoscale gap have attracted great research interest due to hybridization of the localized surface plasmon resonance(LSPR) of individual nanoparticle that are characterized by novel enhanced optical responses.[1, 2]. Because of their tunable plasmonic properties that can be controlled systematically by changing inter-particle feed gap separation, nanoparticle geometry, cavity shape, configuration, composition and refractive index of a medium, it can be utilized in various applications such as sensing, imaging, designing plasmonic devices with tailored optical responses [2, 3]. Optical properties of metallic nanostructures are dominated by coupling of oscillating conduction electrons with electromagnetic radiation field at resonance frequency that gives raise to localized surface plasmon resonance (LSPR) [4-6]. Due to the near field coupling in nanogap between adjacent nanoparticles or hot spot effect, the excitation of localized surface plasmon resonances (LSPRs) in strongly coupled metal nanoparticles can persuade extremely strong localized field intensity enhancement[7, 8]. Thus, it is possible to achieve remarkable plasmonic properties like enhanced optical fields, hybridized plasmon modes[1], and sensitive LSPR shift[9], by adjusting the hot spot configuration or inter-particle distance. Therefore, it can be used in allowing a wide range of applications in surface-plasmon enhanced spectroscopies, photo-catalysis, and optical sensing[10]. Furthermore, the high sensitivity of LSPRs to nanoparticle's shape, size, material composition, and the refractive index of the embedded medium makes it a possible candidate to tailor their optical properties for desired application [4, 5, 11].

On the other hand, the gap distance between two nanoparticles and their nanogap morphology in particular, modifies tunable optical response by affecting plasmon coupling that lead to enhanced near field intensity, formation of new hybridized modes that exhibits unique spectral feature[8]. In our study, the gap distance between the two nanoparticles is a key parameter to understand the behaviour of different plasmon modes emanating from coupling nanoparticles. Various studies have been conducted so far and reported on the emergence of optical property that are dominated by bonding dipole plasmon (BDP) mode in capacitive coupling of nanoparticle dimers[12], charge transfer plasmon(CTP) mode in conductively bridged symmetric and asymmetric nanoparticle dimers[12, 13], Fano resonance(FR) mode in asymmetric dimer[5]. Also it has been reported that an outstanding change of the plasmon modes are predicted when the coupling of the nanoparticle dimer changes to overlapping regimes[14]. However, there is still a lack of comprehensive study that has been carried out to provide brief description in bonding of dipole plasmon of individual nanoparticle hybridized to form bonding dipole plasmon (BDP) modes at lower energy that arise in overlapping asymmetric nanoparticle in a dimers for sensing applications. Plasmonic nano dimers near touching and overlapping limits exhibit a significant shift in plasmon resonance, evolving modes, and disappearing of higher order multipolar modes[15], due to a drastic enhancement of the induced electric field[14].

Moreover, the optical properties of plasmonic nanostructure of on sub-nanometer scale gap separation between two symmetric nano particle dimer configuration until reaching to quantum tunnelling have been reported in previous studies as well by using classical electromagnetic theory based on gap separation scale limit prior to touching [1, 16-19]. When two metallic nanoparticles are positioned in close alignment to one another, symmetrical dimers create an in-phase mode that is optically radiant, while the out-of-phase mode is dark that is optically non-radiant because of their corresponding dipole moments cancel out[5, 20]. More versatile and tunable sensing platforms may be made possible by the asymmetry's potential to produce novel plasmonic modes and coupling effects[20, 21]. It can exhibits collective plasmon resonance modes by systematically controlling the gap separation between individual nanoparticle in a dimer and the cavity shape[8], that can introduce new effects which are not present in the symmetrical configurations [5, 12, 22]. Furthermore, the effect of symmetry breaking in, shape, and composition can introduce a new spectral feature shifting for a linear symmetric homo dimer. In addition to this, cavity shape of constituent nanoparticles plays crucial role in determining spectral response and spatial distributions of surface plasmon resonance[23, 24], hence the precise control and manipulation of dimer nanogap morphology intensely modifies optical properties of dimer system[25]. Therefore, by bringing the two asymmetric nanoparticles in touching and overlapping regime drastically modifies the optical properties of the proposed nano structure by offering additional possibilities for modulating the plasmonic response and qualifies a variety of potential applications such as for sensing, photonics, energy, surface enhanced Raman spectroscopy and so on [5, 19, 26].

To this end, it is important to examine the properties of appearing bonding dipolar plasmon modes of the nanoparticle dimer in touching and overlapping regime by introducing symmetry breaking of the involved nanoparticle in the dimer. Therefore, we systematically investigate the tunable plasmonic properties of spatially overlapping asymmetric nanoparticle dimers using rigorous computational approach; we found significant spectral variation of induced resonance modes that are dominant bonding dimer plasmon resonance (BDP) in two distinct regimes of overlapping symmetric AuND, shape-asymmetric AuND-NP and compositionally asymmetric Au-AgND dimers.

Additionally, we have examined the behaviour of BDP mode of the dimer under variation of the particle shape, nanoparticle gap morphology of the constituent nanoparticle in a dimer and the refractive index of the medium. Furthermore, in view of actual applications in surface-based sensing, the implications of resonance shift of the plasmon mode of the AuND, AuND-NP, and Au-AgND dimers are extensively investigated and it was found that overlapping shape and composition asymmetric nanoparticle dimers shows significantly a larger resonance shift and figure of merit (FOM) of a proposed dimer structure that enables a potential for sensing application.

## 2. Design and Simulation

The studied geometries of symmetric AuND, shape asymmetric AuND-NP, and compositionally asymmetric Au-Ag ND dimer structures are depicted in Fig-1. In order to study the spectral variation in two gap separations, s = 0 and s = -2nm, we first used a symmetrical Au ND dimer, which is composed of two identical nanodisk in Fig. 1(a), each having an equal diameter and thickness and gap separation s = 0nm, -1nm, -2nm. Later, we purposefully mismatched the constituent nanoparticles of the dimer in shape and composition as shown in Fig-1(b) and (c) with equal diameter and thickness in order to investigate the impact of introducing symmetry breaks on the optical responses of the asymmetric dimer at two regimes (s = 0nm and s = -2nm), known as conductively touching and overlapping regimes. Additionally, to extend our investigation by tuning the gap morphology of the dimer, we designed three configurations of gold nanoplate (AuNP) dimers with equal diameter and thickness for the study of the effect of gap morphology in overlapped nanoparticle dimers (see Figs. 1(g)–(i). Lastly, the three dimers were placed on a substrate whose refractive index was changed from 1.31 to 1.36, as shown in Fig. 1 (d)-(f), in order to study the sensing applications of BDP resonance modes.

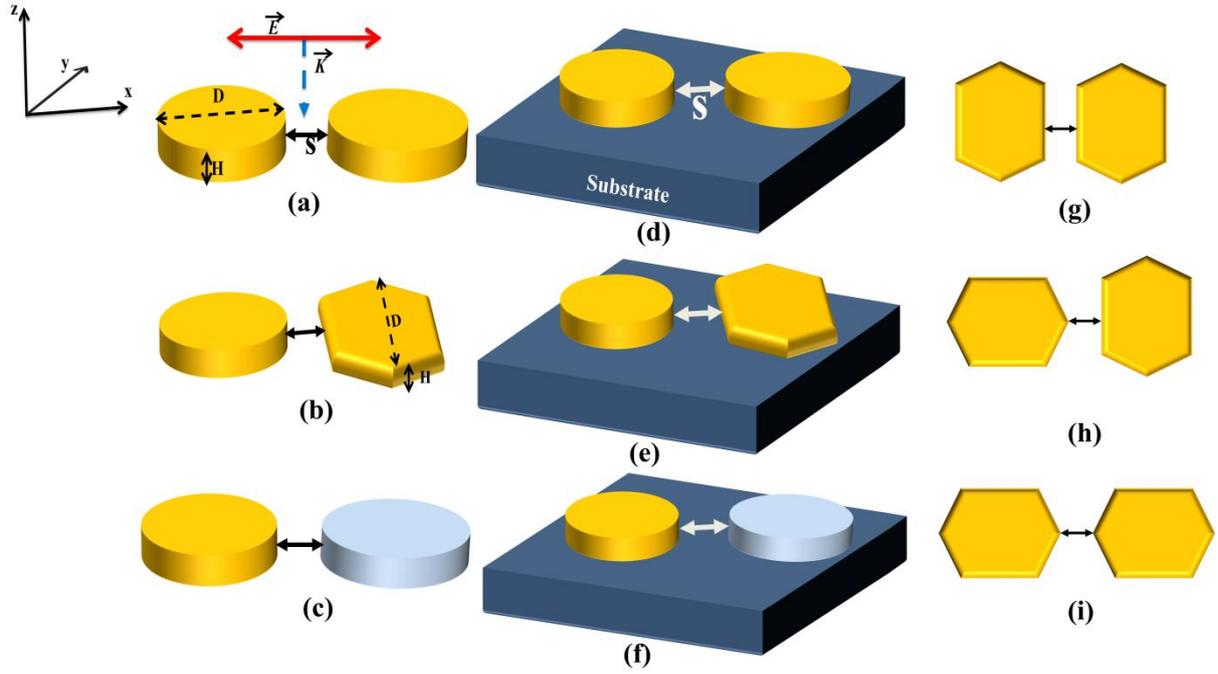

**Fig-1.** Schematic representation of the studied nanoparticle dimers. (a), (b), (c) respectively showing symmetric Au ND, shape asymmetric Au ND-NP, and compositionally asymmetric Au-Ag ND dimers separated by gap distance s, where s = 0 nm, -1nm, and -2nm. (d), (e) and (f) represents symmetric Au ND, shape asymmetric Au ND-NP and compositionally asymmetric Au-Ag ND dimers respectively placed on substrate.(g), (h) and (i) represents changing gap morphology of AuNP. Negative values of s correspond to overlapping nanoparticle dimer.

The diameter D and the height H of the studied structure were D = 100nm, and H = 50nm. The arrows in the inset (a), illustrate the directions of the polarization (red arrow with electric field vector) and wave vector (black arrow).

The dimers were irradiated by a linearly polarized plane wave, which was injected along the z-axis, perpendicular to the plane of the nanodimer systems. The polarization of the incident light was oriented along the dimer axis, as seen in Fig-1(a), The finite-difference time-domain approach was used to calculate the optical response of the dimer systems[27, 28]. It enables to calculate the optical scattering cross section by discretizing time and space and substituting finite differences for derivatives in the solution of Maxwell's equations[29]. The optical responses of the investigated nanostructure dimers, which span a spectral range from the visible to the mid-infrared (MIR) region (500 nm – 2000 nm), were calculated using a total-field-scattered-field (TFSF) source. The dielectric functions of the designed systems were modeled using the experimental data from Johnson and Christy for gold[30]. We have employed perfectly matched layer (PML) boundary conditions in our simulations to ensure complete absorption of electromagnetic radiation at the simulation boundaries[5, 31]. Throughout the investigation, we employed a mesh size of 2 nm in all x, y, and z dimensions. The computation's convergence test has been performed, and the current study's results show that the error is within a satisfactory range [13, 32, 33]

## 3. Result and Discusion
### 3.1. The effect of nanodimer overlapping on plasmon resonance

LSPR spectra are influenced by various factors such as composition, size, shape, inter-particle distance, and refractive index[34], allowing for tunable plasmon resonance and enhanced electromagnetic field[35]. We have observed a dominant resonance peak in the scattering spectra for the three configurations of dimer shown in Fig. 2(a)-(c) respectively. This resonance peak corresponding to bonding dipole plasmon (BDP) mode or dipole-dipole coupled plasmon mode as the nanoparticle dimer starts to make a conductive contact at a single point of gap separation s = 0nm and increasing conductive overlap to s = -2nm[20, 23, 36]. Furthermore, disappearing in the magnitude of the hybridized higher multipolar dimer mode has been occurred[15].Thus, the appearance of this dominant dipolar plasmon mode in touching or overlapping nanodimer can be further characterized by a singular behaviour with a net charge in each of the dimer particle that is confirmed by the surface charge distribution of dimers upon excitation at the designated wavelengths, which clearly reveals the characteristics of BDP modes as shown in Fig. 2 (m)-(o).

Fig-2 (a) shows scattering spectra calculated by varying dimer gap separation s from touching to conductive overlap for the symmetric AuND dimer. Thus, the symmetric dimer exhibits different spectral resonance peak for s = 0nm, -1nm and s = -2nm that are located at the resonance wave length of 1033nm, 1024nm and 1008nm respectively.When the gap separation changes from touching to overlapping regime, the observed BDP mode tends to blue shift with increasing overlap [14, 15, 20]. As a result, we found a resonant peak shift ($\Delta\lambda$ = 25nm) from 1033nm to 1008nm. Thus, blue shifting mode induced by overlapping symmetric dimers is because of a decreased coupling of the individual nanoparticle plasmons, due to the onset of electron transfer between the particle[15]. Moreover, the exact control over the plasmonic features of the nanoparticles is made possible by the peak blue shifting of the dipolar plasmon mode in overlapping symmetric dimers[26]. Thus, it can be used for applications like better

light-matter interactions, more effective energy transfer processes, and improved molecular sensing[21].

Quantitative data of the resonance peak extracted from Fig. 2(a) that is computed and displayed in Fig. 2(d) further supports this blue shifting of the mode linearly as the gap separation increases toward negative (increasing overlap). To further support our result with qualitative data, we have calculated the local electric field distribution of the touching and overlapping symmetric dimer shown in Fig. 2(g). From this result, we can see that, as two nanoparticles move from touching to overlapping regime, the local electric field confinement typically increases, the electric field confinement is limited to the region between the nanoparticles while they are in the touching regime, however due to plasmon coupling effect, the interaction between nanoparticles leads to increased local field enhancement as nanoparticle changed to overlapping regime (see Fig. 2g (i) and (ii)). On the other hand, in overlapping symmetric nanoparticle dimers, relatively decreased local field confinement have been observed compared to overlapping asymmetric nanoparticle dimers presented in Fig. 2(h) and 2(i).

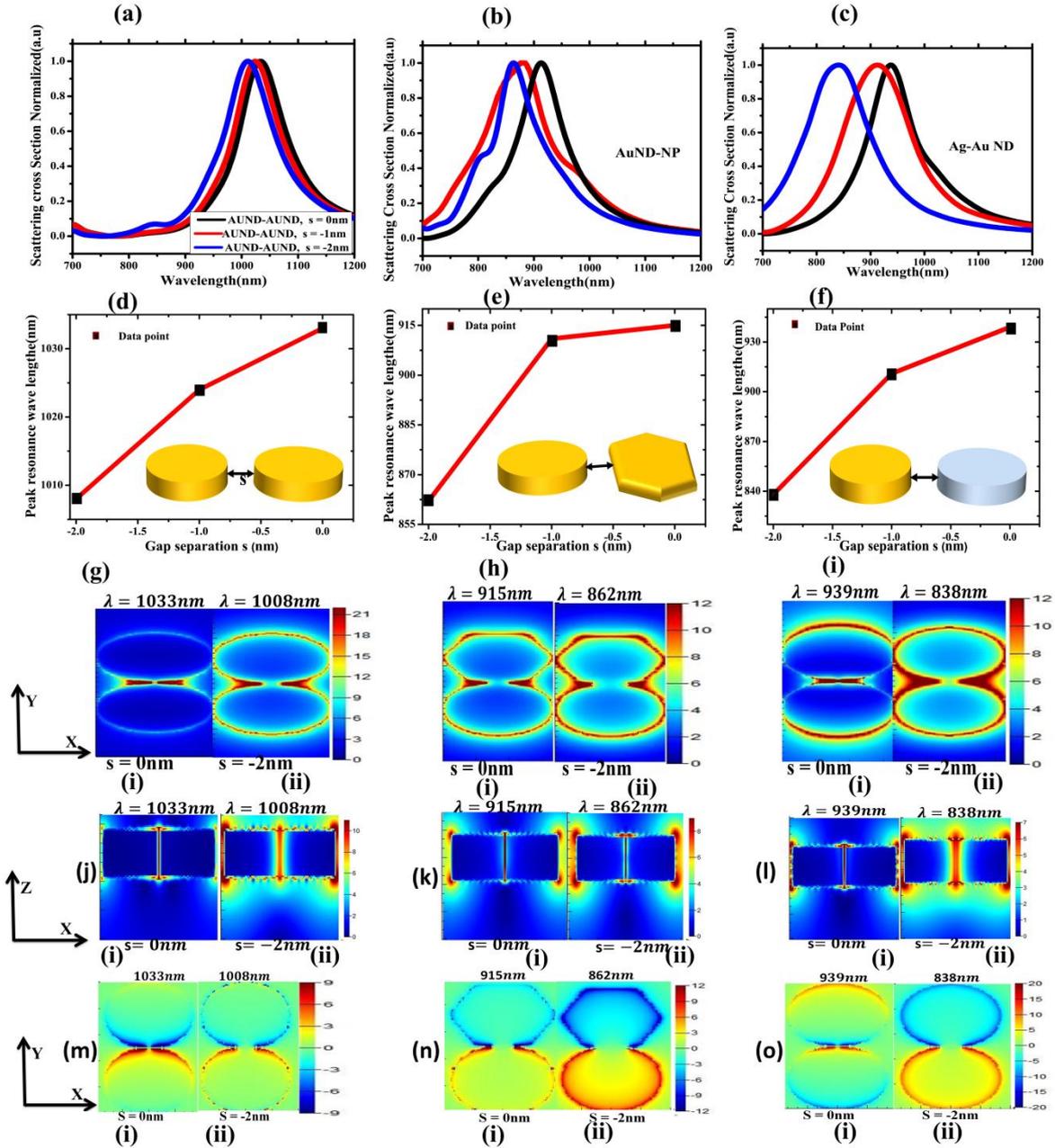

**Fig-2.** (a), (b) and (c) respectively in the first row displays calculated scattering cross section for the symmetric Au ND, shape asymmetric AuND-AuNP and compositionally asymmetric Au-AgND dimer as a function of gap separation (s) for touching (s = 0nm) and overlapping regimes. (d), (e) and (f) respectively in the second row shows the corresponding resonance peak position as function of gap separation s. (g), (h) and (i) respectively in the third row shows the near electric field distribution in XY-cross section for the dimer at two regimes (s = 0nm, and s = -2nm). (j) - (l) similarly shows the near electric field distribution in XZ-dimension for the dimer at two regimes (s = 0nm, and s = -2nm). (m) - (o) presents surface charge distributions

profile corresponding to selected gap separation(s = 0nm, and s = -2nm) in three configuration with particular wavelength points.

Asymmetric nanoparticle dimers exhibit altered optical characteristics and nanoscale electromagnetic interactions upon overlapping, modifying collective optical response by varying composition, shape, alignment, and distance from one another.[37]. The study reveals that morphing nanostructure significantly influences the optical responses of single and coupled nanoparticles. [26]. To investigate its effect, we have designed shape asymmetric Au ND-NP heterodimer, which were kept at a gap distance s, as shown in Fig-1(b).The corresponding scattering spectrum of shape asymmetric Au-ND-NP dimer has been obtained and displayed in Fig-2(b). When the two constituent shape asymmetric gold nano particles are placed at varied inter-particle gap separation of s = 0nm, s = -1nm, and s = -2nm respectively, we have observed a resonance peak at a wave length's points of 915nm, 881nm, and 862nm that shifts to ward high frequency regime, this is because of bonding plasmon modes, where charges are dispersed over both overlapping gold nanoparticles are formed as a result of the hybridization of plasmon modes[38].This suggests that, when the overlapping of the shape asymmetric nanoparticle increases the mode exhibits evident blue shifts that imply the components' energy levels are farther apart, which reduces their energy overlap resulting in a weaker dimer coupling strength[15].

In the same manner as we discussed in overlapping symmetric gold nanoparticle dimer, we observed the spectral peak resonance shift ($\Delta\lambda$ = 53nm) of overlapping shape asymmetric gold nanoparticle dimers as the inter-particle separation of the dimer system changes from touching to overlapping regime (from s = 0nm to s = -2nm). In comparison to spatially overlapped symmetric dimer, the overlapping of nanoparticles in a shape-asymmetric dimer produce a greater resonance shift. This spectral shift can have significant implications for sensing applications when the dimer is placed on a substrate that varying in refractive index. The coupling between the plasmonic modes of the individual nanoparticles is responsible for this amplified shift[26], which results in a stronger interaction and alters the total plasmonic response for various LSPR based applications. In addition, while overlapping shape asymmetric nanoparticle dimers offer more tunable and versatile feature in sensing applications, whereas overlapping symmetric nanoparticle dimers offer well-defined plasmonic resonances and sensitivity[33]. Thus, the quantitative result extracted from Fig-2(b) for overlapping shape asymmetric AuNP dimer reveals the linear relationship between peak resonance wave length and gap separation of the nanoparticle dimers were presented in Fig-2(e) that evidently supports aforementioned discussion. To further elucidate, we have calculated the qualitative results of the corresponding near field distribution as shown in Fig. 2(h). Thus, the result indicates comparatively more local field is confined near to touching region of dimer than near to overlapping region of dimer (See Fig. 2(h) (i)-(ii). But if we compare it with symmetric dimer, the local field confinement is more pronounced in shape asymmetric dimer. Therefore, in general the higher local field confinement related with increased field intensity[7] that can lead to spectral blue shifting of plasmon resonance associated with nanoparticle.

We separately calculated the scattering spectrum of a compositionally asymmetric heterodimer composed of AuND and AgND as shown in Fig-1(c) to further elucidate compositional effect in overlapping dimers. This allows us to show the impact of introducing compositional asymmetry on the plasmon coupling phenomenon[5]. As illustrated in Fig-2 (c), by varying the gap separation s of the dimer as s = 0 nm, -1 nm, and -2 nm respectively, we have found resonance peak in wave length point of 939 nm, 911 nm, and 838 nm. Therefore, the observed peak spectral shift toward higher frequency regime in the of Au–AgND heterodimer configuration has significant implications for plasmonic applications[9]..

Moreover, as we can see from the result, the resulting mode abruptly blue shifts with a higher resonance shift ($\Delta\lambda$ = 101nm) as compared to symmetric and shape-asymmetric dimers as the two constituent nanoparticles turns from touching to overlapping regime of the dimer. This enhanced blue shift in the Au-AgND dimer is because of the dominance of the silver nanoparticle in the dimer. The plasmon resonance peak of nanoparticles made of silver shows resonance peak toward blue side where as gold nanoparticle with equal size display their plasmon resonance peaks toward red side[39]. So, compared with the symmetric Au-AuND and AuND-NP homodimer, compositionally asymmetric Au-AgND heterodimer peak position is expected to give more peak shift. To further confirm, the quantitative result extracted from Fig-2(c) for compositional asymmetry revealing the linear relationship between peak resonance wave length and gap separation of the nanoparticle dimers were presented in Fig-2(f) that evidently supports presented result. On the other hand, we calculated the local field distribution and found that higher local field confinement has been observed near to the overlapping region than touching region (See Fig. 2(i), (i)-(ii). This indicates it has direct effect to increase intensity of electromagnetic field in the confined region[7], resulting a greater spectral shift in magnitude compared to the situation with less local field confinement[40]. Moreover, dimers exhibiting small overlap regime generate a greater local field enhancement compared to nearby non-touching particles, indicating a potential approach to increase the degree of enhancement in applications like SERS[14].

Additionally, cavity shape, the spatial configuration of the gap between nanoparticles in a dimer structure, significantly influences the plasmonic resonance behavior and near-field enhancement of metallic nanostructures [8, 25] . Therefore, one can optimize the near field enhancement by controlling the shape of the cavity in plasmonic nanoparticle dimers through precise engineering of nanoparticle size, shape, and orientation[3, 25]. In order to investigate this effect in XZ view, we have calculated the near field distribution of symmetric AuND, shape asymmetric AuND-NP and compositionally asymmetric Au-AgND overlapping dimers in XZ-dimension and presented in Fig-2 (j) - (l). From this result we can see that near field distribution has enhanced in uniform gap geometry as seen along Z-dirction of touching and overlapping region of AuND, Au-AgND dimers shown in Fig-2 (j) and (l)[25]. However in the XZ view of Au ND-NP as shown in Fig-2 (k), in contrast to the electric field distribution of other configuration, the electric field near a gold nanoplate corner is more dispersed and intensely concentrated at the surface in overlapping nanoparticle dimer. This can be explained by the gap geometry difference and the one-dimensional overlap[41].

Both shape- and compositionally-asymmetric nanodimers, which consist of two different shapes and compositions of nanoparticles, provide different plasmonic responses than their symmetric counterparts[42]. These nanoparticles are very attractive for many applications, such as refractive index sensing, surface plasmon-enhanced spectroscopies, and photocatalysis[5]. The precise change in the resonance spectral peak depends on various elements, including the nanoparticles' geometry, composition, and inter-particle distance, as well as the refractive index of the surrounding medium. Different spectral resonance shifts can be achieved in overlapping asymmetric nanoparticle dimers by tuning these parameters[9]. As we can see from the results displayed in Figs. 2(b) and (c), the peak resonance shift of overlapping shape and composition asymmetric nanoparticle dimers gives significantly a larger resonance shift, which is $\Delta\lambda = 53$nm and $\Delta\lambda = 101$ nm, respectively, compared to the resonance shift of overlapping symmetric nanoparticle dimers, which is $\Delta\lambda = 25$nm. Not only shape or gap induced, but also this enhanced shift can be attributed to the intrinsic properties of the composite materials. Furthermore, overlapping asymmetric dimers have higher sensitivities than their symmetric counterparts[43]. Because the individual nanoparticles in asymmetric dimers have distinct shapes and compositions, they may exhibit tunable plasmonic resonances. This makes it possible to tailor the plasmonic properties to fit certain applications and to provide a broader range of spectral responses. Thus, plasmonic responses can be tuned more freely when the shape or composition of the dimers is present, which increases sensitivity in sensing applications[43]. For sensing applications, the dimer configuration with a greater resonance shift can be suggested as a viable option and employed extensively[33].

### 3.2. Effect of gap morphology in overlapped nanoparticle dimer.

The plasmonic properties of a nanoparticle dimer and the accompanying optical phenomena can be strongly influenced by the gap's morphology[41]. The gap geometry within a nanoparticle dimer significantly affects the plasmon resonance mode, its near-field interactions, electromagnetic field enhancement, and spectral properties. Therefore, understanding and controlling the gap geometry is vital in tailoring the optical characteristics and uses of nanoparticle dimers in various fields of application like photonic devices, sensing, and imaging. In particular to our dimer configuration under the study, it would be interesting to represent quantitatively and qualitatively how the shapes of individual nanoparticles in the dimer at nearly contact (touching and overlapping) regions influence the resulting optical properties of the nanodimer, such as plasmon resonance shift and the near field enhancement. To this end, we have presented AuNP dimers with three different configurations with changing shapes in the contact region that have been kept at a gap separation of s, where s = 0nm, -1 nm, and -2 nm, as shown in Fig. 3(a)–(c). Our calculated results in Fig. 3(a), (b), and (c) respectively elucidate three spectral peak wavelengths of scattering cross-section spectra at 777nm, 771 nm, and 759nm for edge-to-edge contact of the AuNP dimer; 1026nm, 977 nm, and 891nm for edge-to-vertex contact of the AuNP dimer; 1125 nm, 992 nm, and 773 nm for vertex-to-vertex contact of the AuNP dimer as a function of three gap separations s = 0nm, -1 nm, and -2 nm that obviously shifts as increasing overlap of constituent nanoparticles in the dimer. Therefore, the resonance shift for edge to edge, edge to vertex, and vertex to vertex overlapped dimers is $\Delta\lambda = 18$nm, $\Delta\lambda = 135$ nm, and $\Delta\lambda = 352$ nm, respectively. By altering the gap morphology, we have obtained significant variation in the plasmon resonance shift, which influences the coupling between the nanoparticles, leading to changes in the overall optical response of the dimer. In comparison among three dimers, we have found that the resonance shift of the overlapping vertex-to-vertex dimer configuration in Fig. 3(c) is significantly red-shifted in the resonance peak (greater resonance shift), which magnifies the fact that a more confined gap between the nanoparticles due to the sharp contact vertex may result in increased electromagnetic field enhancement and stronger plasmon coupling than that of an edge and a flat touching surface[23]. This may lead to more noticeable alterations in the dimer's optical response and a greater resonance shift as nanogap morphology changes.

In general, the difference in the gap morphology between dimers in contact region can significantly impact on the coupled plasmons resonance mode and field enhancement and further suggest exciting applications in plasmonic sensing or surface-enhanced spectroscopies[44]. To further support our result, similarly we have calculated corresponding local field distributions of three dimers at resonance peak wave length extracted from the scattering cross section spectra are presented in Fig-3(d),(e) and (f). It should be noticed that the confined and enhanced local field of vertex to vertex overlapped AuNP dimer shown in Fig. 3(f) (i) - (ii) among other dimer configuration result in a significant increase in the intensity of electromagnetic field at the location[7]. This increased intensity can lead to higher spectral blue shifting of the plasmon resonance associated with the nanoparticles dimer (See Fig. 3(c)).Thus, surface-enhanced spectroscopies (such as surface-enhanced Raman spectroscopy), nonlinear optics, and sensing are among the many applications for which the relationship between local field confinement and spectral blue shifting is significant[9]. This suggests that, in vertex to vertex overlapping, the constituent nanoparticles interact as seen by the strong coupling of their respective near fields than edge to edge and edge to vertex overlapped AuNP dimer configuration.

In addition, to observe cavity shape effect that influences the plasmonic mode excited in a dimer and the near field enhancement distribution with in the gap region of touching and overlapping AuNP dimer[25]. We have calculated the near field distribution in XZ-dimension in three configuration of AuNP dimer in similar manner and presented in Fig-3(g), (h) and (i). . From our result, we can see that more near field is confined in a gap region of dimers with sharp vertex of contact (see Fig-3(h) and (i)) because it create a smaller and more confined gaps between nanoparticles that leads to strong plasmon coupling and higher electromagnetic field enhancement with in a gap as viewed in XZ-dimension. As a result of this, larger resonance shift have observed comparatively in AuNP dimer with vertex to vertex touched and overlapped region. The higher near field enhancement viewed in AuNP dimer with edge to vertex configuration than that of edge to edge configuration is due to sharp vertex effect of one of constituent particles in a dimer. Therefore, the differences in cavity shape between dimers can significantly influence the plasmon resonance behaviour and optical properties[23, 25].

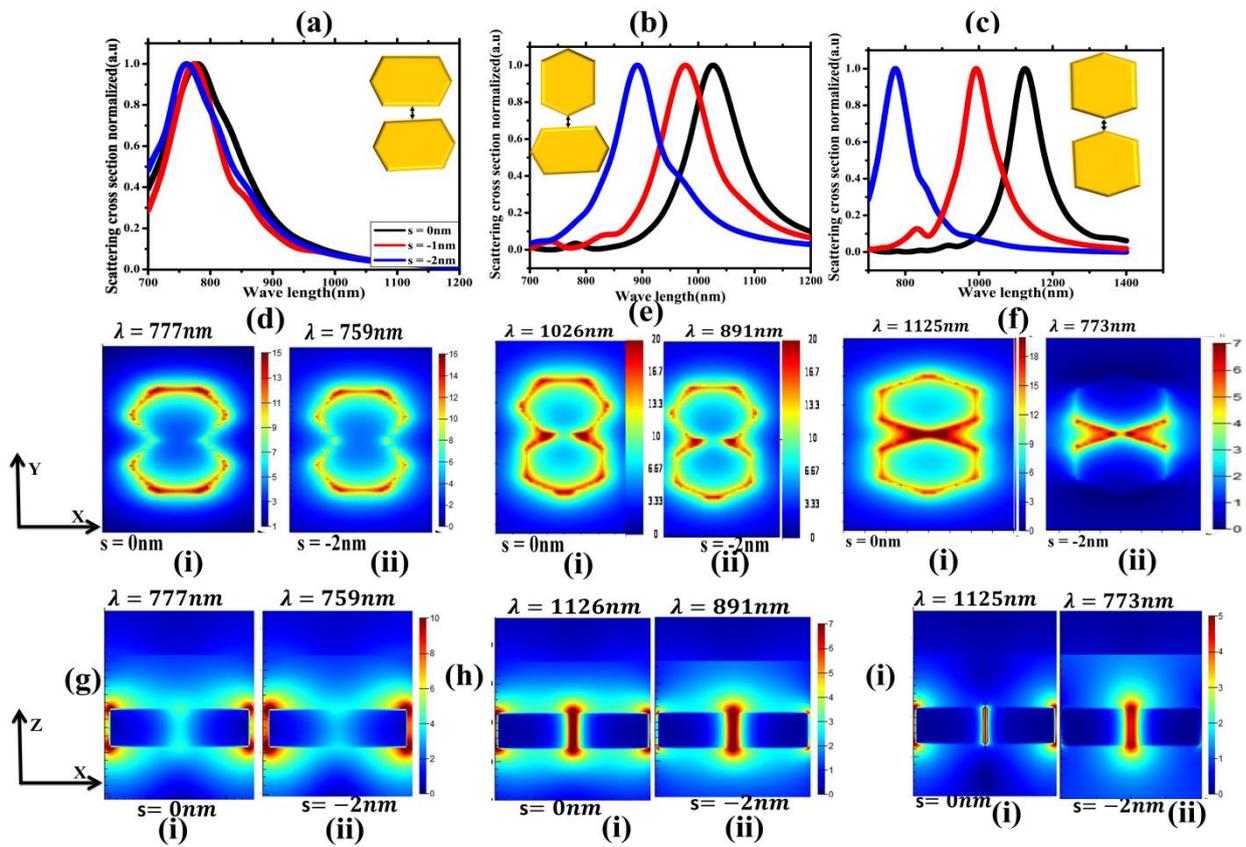

Fig-3. (a), (b) and (c) respectively presents the scattering cross section of the three configuration of dimers with different gap morphology(shown as inset) as a function of gap separation s. (d), (e) and (f) respectively reveals the near field distribution in XY direction for the dimer at two regimes (s = 0nm, and s = -2nm). (g), (h) and (i) similarly shows local field distribution in XZ dimension for three configuration of dimer based on their gap morphology at two regimes(s = 0nm, and s = -2nm). The inset in (a),(b) and (c) respectively represents the geometries with varying gap morphology.

### 3.3. Effect of thickness in overlapped nanoparticle dimer

The thickness of the nanoparticle dimer is the most common and important geometrical parameter that significantly affects plasmonic coupling and electromagnetic interactions within the dimer structure[45, 46]. Thus, the plasmon resonance shift in overlapping nanoparticle dimers can be substantially modified by the thickness of the individual nanoparticles [45]. To investigate the effect of this parameter, we have calculated the scattering cross section of the symmetric gold nanodisk (AuND) and gold nanoplate (AuNP) dimer configurations that are presented in Figs-4(a)–(d). From this, we can observe, respectively, a significant blue-shift (and red shift) of the bonding plasmon resonance (BDP) mode that results from increasing (and decreasing) the thickness of the nanoparticle by +2nm and -2nm from our reference thickness H = 50nm in a dimer [46]. This indicates the thickness of AuND and AuNP are very sensitive parameters to change the plasmonic properties of the dimer system.

However, the degree of shifting varies between the two separate regimes of the dimer, s = 0 nm and s = -2 nm in both the symmetric AuND and symmetric AuNP dimers. Resonance peak shifts of $\Delta\lambda$ = 4 and 7 nm respectively, in an AuND-ND dimer with gap separation of s = 0 nm indicate blue and red shifting in spectral peak. In contrast, a remarkable increase in resonance peak shift of $\Delta\lambda$ = 10 and 11 nm in the same dimer at gap separation of s = -2 nm, describes blue and red shifting in spectral peak wave length. In the same manner, we can see the resonance peak shift for AuND-NP at gap separation s, where s = 0nm and -2nm. As a result, the resonance peak shift $\Delta\lambda$ = 5 nm and 8 nm, respectively, for s = 0 nm that indicates blue and red shifting in the spectral peak wave length. The resonant peak shift for s = -2nm in similar dimer AuNP-NP is $\Delta\lambda$ = 15nm and 25nm respectively, shows significantly higher in resonance shift. From this, we can conclude that the degree of resonance shift varies in terms of dimer configuration; the AuNP dimer manifests a higher shift than that of the AuND dimer (See Figs-4(a)–(d)).

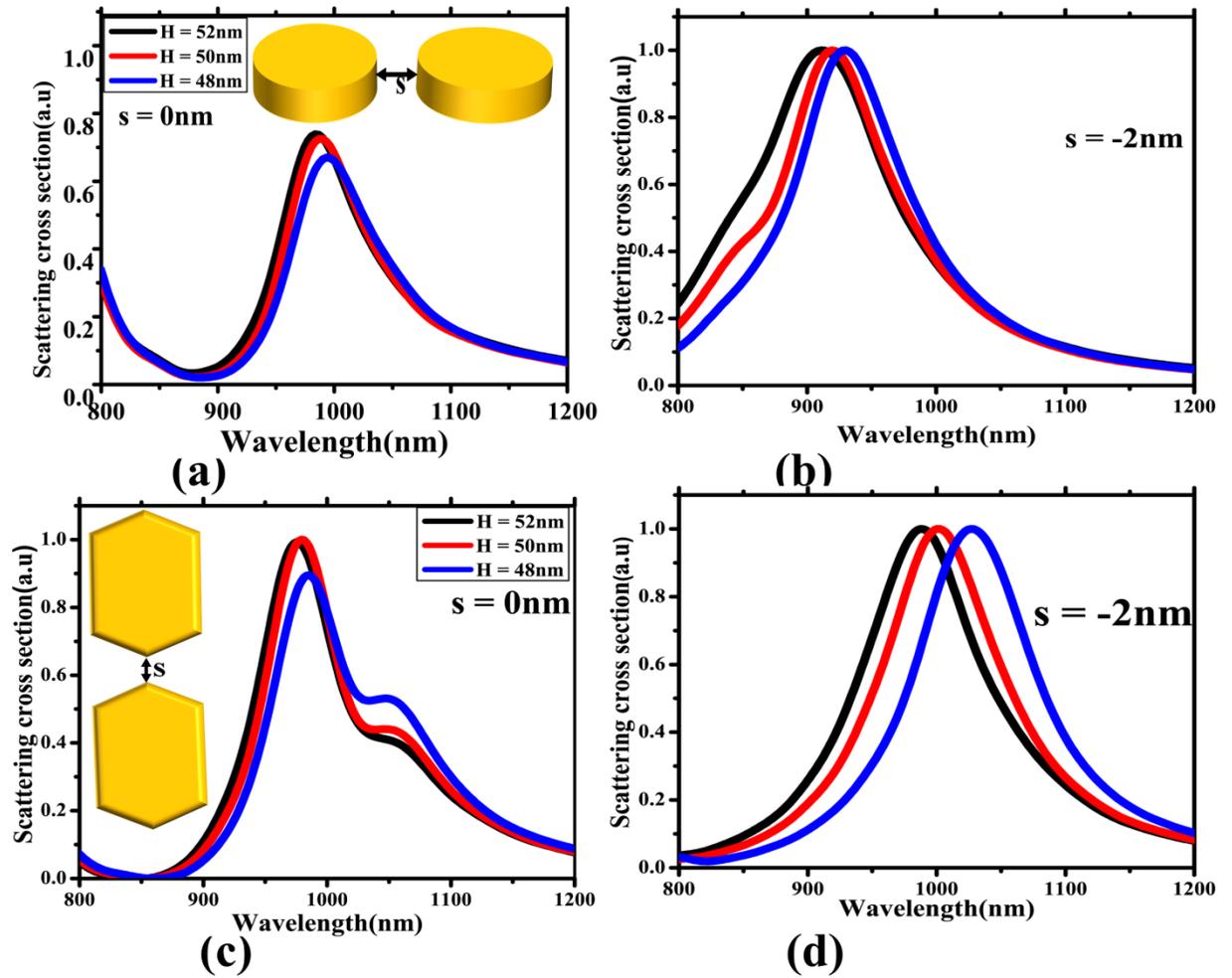

**Fig-4.** (a) - (d) represents scattering cross section as a function of thickness of the symmetric AuND and AuNP dimer of gap separation s at touching (s = 0nm) and overlapping (s = -2nm) regimes respectively.

### 3.4. Sensing application of bonding dimer plasmon modes in overlapped nanoparticle dimers

The dielectric environment is another critical factor affecting LSPR coupling[43]. The shape, composition, and structures of nanoparticles all have a significant impact on plasmonic sensitivity[47, 48]; therefore, plasmonic nanoparticles are the suitable platforms for refractive index measurement. The most important parameters to evaluate the sensor performance are the sensitivity and figure of merit (FOM)[13, 49].

This section examines the sensitivity of bonding dimer plasmon modes to change in the ambient refractive index of the substrate where overlapping symmetric, shape-asymmetric, and compositionally-asymmetric nanoparticle dimers are placed on it. We used optimized spectra to find the scattering spectrum's peak position as a function of the refractive index. This helped us figure out the sensor's sensitivity and figure of merit (FOM), as shown in Fig. 5(a), (c), and (e).

As shown with black curves in Fig-5(a), (c), and (f), the resonance wavelengths of the BDP modes in overlapping symmetric, shape-asymmetric, and compositionally asymmetric nanoparticle dimers with a refractive index of 1.31 are 970 nm, 870 nm, and 980 nm, respectively. When the refractive index increases from 1.31 to 1.36, this mode shows a red shift from 970 nm to 982 nm, 870 nm to 885 nm, and 980 nm to 999 nm, respectively as shown by the blue curve in Figs. 5(a), (c), and (e). The shift of the ambient refractive index from 1.31 to 1.36 results in a shift of the resonance wavelength by 12 nm, 15 nm, and 19 nm, respectively (see Fig. 5(b), (d), and (f)), which is consistent with previous reports[21, 50]. With increasing refractive index, it is evident from Figure 5(a), (b), and (e) that the BDP mode is more red-shifted in overlapping asymmetric nanoparticle dimers than in overlapping symmetric nanoparticle dimers.

In particular, for plasmonic sensors, the shift in the wavelength of the plasmon resonance peak with respect to variations in the refractive index of the surrounding medium ($\Delta\lambda/\Delta n$) could provide important information[13, 51]. In addition, here the change of resonance wavelength with respect to variations of refractive index can be used to quantify the sensitivity (S) of the BDP mode of a plasmonic sensor ($S_{BDP}$= ($\Delta\lambda/\Delta n$), which is extremely dependent on the refractive index of the surrounding medium[5, 7, 33], therefore, enhancing the resonance spectral shift can increase the sensitivity[9]. Therefore, as shown in Fig. 4(b),(d), and(f), the sensitivity of BDP modes in overlapping symmetric, shape asymmetric, and compositionally asymmetric nanoparticle dimers is 240nm$(RIU)^{-1}$, 300nm$(RIU)^{-1}$, and 380 nm$(RIU)^{-1}$, respectively. This result supports the notion that overlapping asymmetric nanoparticle dimers are more sensitive than their symmetric counterparts. Moreover, overlapping compositionally asymmetric nanoparticle dimers showed higher sensitivities than overlapping shape-asymmetric nanoparticle dimers. The main reason for these higher sensitivities is the large shift in BDP modes that happens when the refractive index of the substrate changes from 1.31 to 1.36.

FoM is a key factor often used to evaluate the performance of BDP modes in overlapping symmetric, shape-asymmetric, and compositionally asymmetric nanoparticle dimers[12]. The efficiency of BDP in overlapping asymmetric dimers is usually evaluated by their FoM, which is defined as the ratio of sensitivity to resonant linewidth (FWHM: Full Width at Half Maximum), or FOM = $(\frac{S_{BDP}}{FWHM})$ [9, 52, 53]. A larger FOM represents better performance and thus serves as a measure of how "good" a sensor is[33]. As shown in Fig-5 (b), (d), and (f), the values of FOM for the overlapping symmetric, shape asymmetric, and compositionally asymmetric nanoparticle dimers were determined to be 1.55, 2.08, and 3.04, respectively. This result reveals that the value of FOM for compositionally asymmetric overlapping nanoparticle dimers is higher than that of a shape-asymmetric overlapping nanoparticle dimer and its symmetric counterparts, indicating the significant tunability of the BDP resonance mode by the refractive index change of the substrate. Thus, comparatively among the three dimer configurations, the highest value of FOM demonstrates great sensing performance, showing its potential applications as chemical and biological sensors [33, 54].

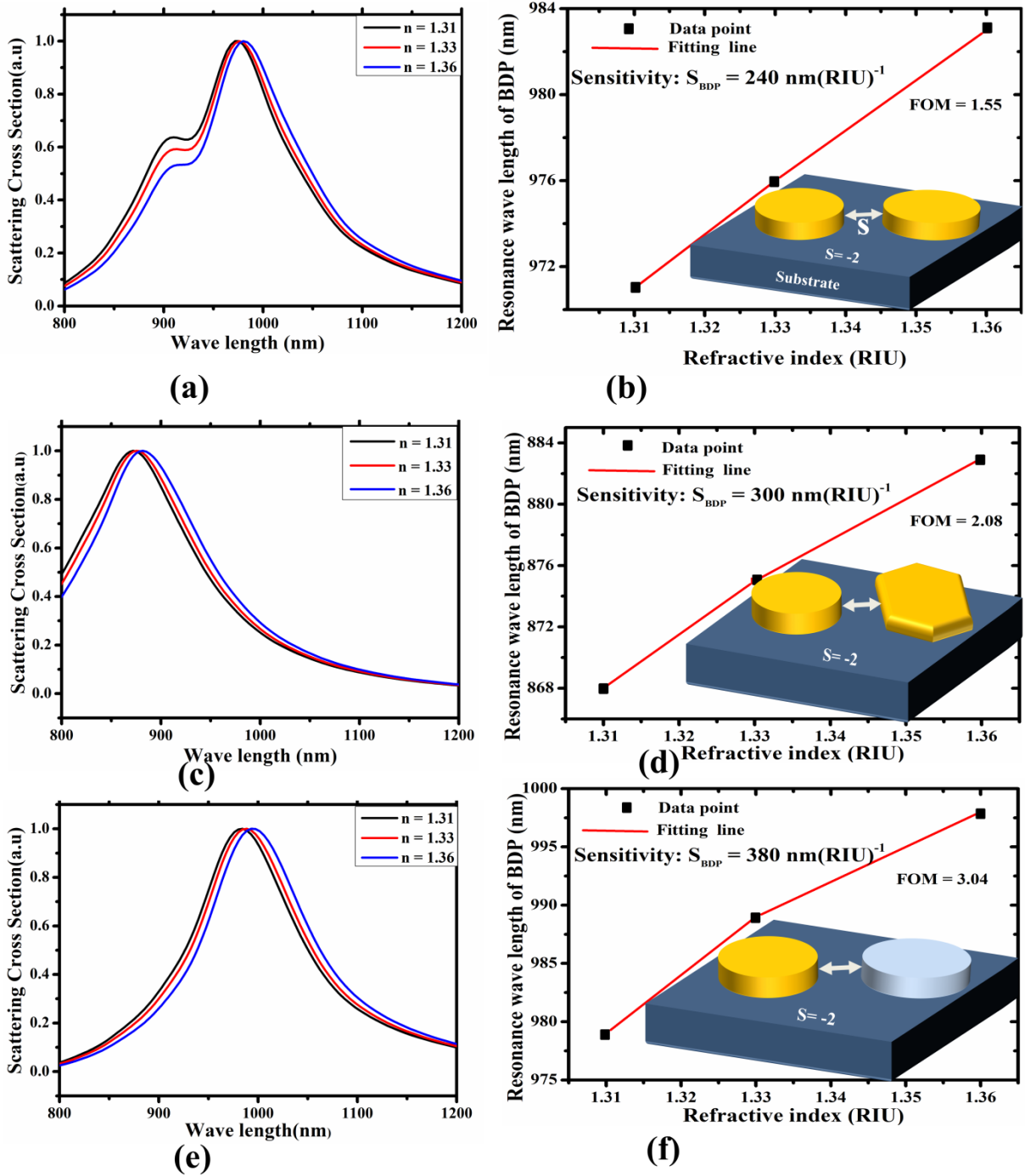

**Fig-5.** Sensitivity of overlapping symmetric and asymmetric nanoparticle dimers as a function of the refractive index of the substrate. (**a**), (**c**) and (**e**) show the calculated scattering spectra of spatially overlapping symmetric, shape asymmetric and compositionally asymmetric nanoparticle dimers respectively with fixed gap separation (s = -2 nm) as a function of the refractive indices. (**b**), (**d**) and (**f**) are the linear plot of BDP peak shifts of overlapping symmetric, shape asymmetric and compositionally asymmetric nanoparticle dimers respectively as a function of the refractive indices of the substrate.

## 4. Conclusion

The study explores the plasmonic properties of spatially overlapped dimers made up of two constituent nanoparticles, which can be tuned over a wide spectral range by adjusting the inter-particle gap, nanoparticle geometry, gap morphology, composition, and refractive index. The spatial overlapping of symmetric, shape, and compositionally asymmetric nanoparticle dimers introduces a new plasmonic phenomenon, including the appearance of a dominant BDP mode that spectrally shifts as the gap separation changes from touching to overlapping. Thus, we have obtained a spectral shift of $\Delta\lambda = 25$nm, $\Delta\lambda = 53$ nm, and $\Delta\lambda = 101$ nm, respectively, for three dimer configurations, and the corresponding advantages of increasing spectral shift were discussed. It has also been found that blue shifting of the plasmon mode occurred by changing the gap morphology of the contacting region of the dimer as two gap separation changes from touching to overlapping, which supports the aforementioned result. In addition, we have found a significant resonance shift by changing the thickness of the overlapping nanoparticle dimer, which is an essential parameter to optimize the performance of nanoparticle-based devices. Finally, we explored the sensing applications of bonding dimer plasmon mode with optimized geometries to evaluate the ability of BDP to detect refractive index sensitivity. Thus, the computed refractive index sensitivity and FOM of the overlapping symmetric, shape-asymmetric, and compositionally asymmetric nanoparticle dimers were 240nm $(RIU)^{-1}$, 300nm $(RIU)^{-1}$, 380 nm $(RIU)^{-1}$ and 1.55, 2.08, 3.04 respectively, and comparatively, the implications for surface-based sensing have been thoroughly discussed.

**Funding**

This project was supported by National key research and development program No. 2022YFA1604304, 2022YFA1604303; The National Natural Science Foundation of China (NSFC) (62005022, 12004052, 62175018, U22A2070), Department of Science and Technology of the Jilin Province (YDZJ202201ZYTS30, 20220508137RC, YDZJ202201ZYTS299, 20200201268JC, 20210402072GH, 20200301042RQ), Department of Education of Jilin Province (JJKH20220720KJ), Jilin Provincial Key Laboratory of Ultrafast and Extreme Ultraviolet Optics (YDZJ202102CXJD028), "111″ Project of China (D17017), and the Ministry of Education Key Laboratory for Cross-Scale Micro and Nano Manufacturing, Changchun University of Science and Technology.